\def\a{\alpha}\def\d{\delta}

\def\l{\lambda}\def\m{\mu}\def\n{\nu}\def
\p{\pi}\def\r{\rho}\def\s{\sigma}
\def\y{\eta}\def\x{\xi}

\def\D{\Delta}\def\L{\Lambda}

\def\ha{{1\over 2}}
\def\app{\approx}

\def\mn{{\mu\nu}}

\def\bh{black hole }
\def\tran{transformations }\def\coo{coordinates }
\def\bg{background }

\def\pb{Poisson brackets }

\def\sch{Schwarzschild }\def\mi{Minkowski }\def\ads{anti-de Sitter }
\def\poi{Poincar\'e }
\def\des{de Sitter }
\def\ades{(anti)-de Sitter }\def\QM{quantum mechanics }

\def\SR{special relativity }

\def\cor{commutation relations }

\def\section#1{\bigskip\noindent{\bf#1}\smallskip}

\def\PL#1{Phys.\ Lett.\ {\bf#1}}\def\CMP#1{Commun.\ Math.\ Phys.\ {\bf#1}}
\def\PRL#1{Phys.\ Rev.\ Lett.\ {\bf#1}}
\def\PR#1{Phys.\ Rev.\ {\bf#1}}\def\CQG#1{Class.\ Quantum Grav.\ {\bf#1}}
\def\GRG#1{Gen.\ Relativ.\ Grav.\ {\bf#1}}
\def\JMP#1{J.\ Math.\ Phys.\ {\bf#1}}

 \def\IJMP#1{Int.\ J. Mod.\ Phys.\ {\bf #1}}
\def\MPL#1{Mod.\ Phys.\ Lett.\ {\bf #1}} 

\def\AoP#1{Ann.\ Phys.\ {\bf#1}}

\def\grq#1{{\tt gr-qc/#1}}\def\arx#1{{\tt arXiv:#1}}

\def\ref#1{\medskip\everypar={\hangindent 2\parindent}#1}
\def\beginref{\begingroup
\bigskip
\centerline{\bf References}
\nobreak\noindent}
\def\endref{\par\endgroup}

\def\up{uncertainty principle }\def\ev{expectation value }
\def\ades{(anti)-de Sitter }\def\QM {quantum mechanics }
\magnification=1200
{\nopagenumbers
\line{\hfil September 2009}
\vskip40pt
\centerline{\bf Extended uncertainty principle and the geometry of \ades space}
\vskip80pt
\centerline{
{\bf S. Mignemi}\footnote{$^\ddagger$}{e-mail:
smignemi@unica.it}}
\vskip10pt
\centerline {Dipartimento di Matematica, Universit\`a di Cagliari}
\centerline{viale Merello 92, 09123 Cagliari, Italy}
\centerline{and INFN, Sezione di Cagliari}
\vskip100pt
\centerline{\bf Abstract}

\vskip10pt
{\noindent
It has been proposed that on \ades background, the Heisenberg \up should be
modified by the introduction of a term proportional to the cosmological constant.
We show that this modification of the \up can be derived straightforwardly from the
geometric properties of \ades spacetime.
We also discuss the connection between the so-called extended generalized \up and
triply special relativity.}
\vskip100pt\
\vfil\eject}

\section{1. Introduction}
The Heisenberg uncertainty principle of quantum mechanics, $\D x_i\D p_j\ge\d_{ij}/2$
plays a crucial role in the conceptual framework of \QM and can be considered one of
cornerstones the of the theory. However, it may be possible that in
extreme situations, far from the range of energy from which \QM was derived, the \up
needs to be modified.

For example, it has been argued that in an \ades background the Heisenberg
uncertainty principle should be modified by introducing corrections
proportional to the cosmological constant $\L=-3/l_H^2$, with $l_H$ the \ades
radius\footnote{$^1$}{In the following, in order to simplify the notations,
we assume $l_H^2<0$ for de Sitter spacetime, and $l_H^2>0$ for \ades.
We use units such that $c=\hbar=1$}, as [1]
$$\D x_i\D p_j\ge{\d_{ij}\over2}\left[1+{(\D x_i)^2\over l_H^2}\right].\eqno(1)$$
This modification has been called extended \up (EUP) [2].
It has been motivated either by analogy with the generalized uncertainty principle
(GUP), which is supposed to hold at quantum gravity scales, and postulates [3-5]
$$\D x_i\D p_j\ge{\d_{ij}\over2}\left[1+l_P^2(\D p_i)^2\right],\eqno(2)$$
with $l_P$ the Planck length, or by gedanken experiments in which the expansion of
the universe during a measurement is taken into account [6].

Although the correction to the Heisenberg formula given by (1) is negligible for
cosmological values of $l_H$, it has a theoretical interest, since it can be used
to derive the correct
value of the temperature of an \ades black hole\footnote{$^2$}{The Hawking
temperature may also be obtained without introducing modifications of the uncertainty
principle, but by using the dynamics of the gravitational field [7].} [1].

The original derivation of (1) was based essentially on analogy with the GUP.
In this letter, we give a more cogent argument, showing that the EUP can be
derived straightforwardly from the geometry of \ades spacetime.
In fact, in order to define quantum mechanics on a
curved background, one must take into account its symmetries. In the case of
\ades spacetime, in particular, the generators of translations do not satisfy
the same algebra as in flat space. It follows from the Jacobi identity that also
the commutation relations between momentum and position \coo are affected.
Since the \up follows from the relation
$$\D x_i\D p_j\ge\ha\left|<[x_i,p_j]>\right|,\eqno(3)$$
where $<\ >$ denotes the \ev, the nontriviality of the \cor implies its
modification.

A straightforward extension of this argument also permits to derive
the more general extended generalized \up (EGUP), obtained by combining (1) and (2),
from a model of doubly \SR (DSR) on an \ades background.

\section{2. Generalized and extended uncertainty principles}
The GUP (2) was first proposed in the context of string theories [3], and then
derived for pure gravity from black hole gedanken experiments [4], non-commutative
geometry [5], or models of measurements that included the effect of the dynamics
of gravitational interactions [8].

For $l_P^2>0$, eq.\ (2) implies the existence of an absolute minimum in the position
uncertainty, $x_{min}=2l_P$.
If one took instead $l_P^2<0$ no lower bound on measurable length would arise, but
rather an upper limit on the momentum attainable by a particle, as in DSR
theories [9]. These theories are based on a definition of spacetime  in which the
action of the \poi group is nonlinearly
deformed in such a way that the Planck energy becomes an observer-independent
constant, which sets an upper limit on the energy-momentum of elementary particles.
Since such deformation is not unique several different models can be defined [10],
starting from these assumptions.

In fact, a derivation of the GUP from a DSR model had been already proposed in ref.\
[5]. Another possibility is to obtain (2) from the Snyder model [11]. The Snyder model
can be interpreted as an example of DSR defined on a noncommutative spacetime, in
which Lorentz invariance is undeformed [12]. The \cor of the Snyder model read
$$[x_\m,x_\n]=i\,l_P^2\,J_\mn,\qquad [x_\m,p_\n]=i\left(\y_\mn+l_P^2\,p_\m p_\n\right),
\eqno(4)$$
where $J_\mn$ are the generators of Lorentz \tran and $\m,\n=0,\dots,3$.
The first equation displays the noncommutativity of the geometry.
It is easy to see that the GUP follows from the second equation, when the \ev
of the right hand side is taken.

As remarked above, the usual DSR interpretation is obtained for $l_P^2<0$.
When $l_P^2>0$, instead, the model has rather different physical properties, like the
existence of a minimal length, that we shall discuss elsewhere [13].
For analogy with \ads spacetime, we call the model with $l_P^2>0$ anti-Snyder.
\bigskip
The GUP can be used to obtain corrections to the heuristic derivation of the Hawking
temperature for \sch black holes [14]. According to the standard argument, one may
identify the uncertainty in the
position of a particle emitted by Hawking effect as $\D x_i\app r_+$, where $r_+$
is the radius of the horizon of the black hole, and its uncertainty in energy as
$\D E\app \D p_i\app 1/(2\D x_i)$, where in the last step we have used Heisenberg
uncertainty principle.
Identifying $\D E$ with the Hawking temperature, one then obtains,
$$T\app{\D E\over2\p}={1\over4\p r_+}\,,\eqno(5)$$
where the numerical factor $1/2\p$ has been introduced in order to obtain the correct
normalization.

If one had instead used the GUP, one should have used for $\D p_i$ the value following
from (2), and hence
$$T\app{\D E\over2\p}={r_+\over8\p l_P^2}\left[1-\sqrt{1-{4l_P^2\over r_+^2}}\right],
\eqno(6)$$
which gives rise to corrections of order $(l_P/r_+)^2$ to the Hawking formula.
If $l_P^2>0$, when the radius of the \bh reaches the minimal length $x_{min}=2l_P$,
the evaporation stops, leaving a remnant [14].

\bigskip
Duality in phase space suggested another generalization of the uncertainty principle
[1], the EUP (1), that should hold on a curved background. In this case,
the length scale $l_H$ can been identified with the \ades radius.
In ref.\ [6] a better motivated derivation of the EUP was given, based on a
gedanken experiment taking into
account the effects due to the expansion of the universe during a measurement.

If $l_H^2>0$ (\ads space), the relation (1) implies the existence of a minimal
value for the momentum of a particle, given by $p_{min}=2/l_H$, while if $l_H^2<0$
(\des space) there is no constraint on the momentum, but a maximum length emerges,
given of course by the radius $l_H$ of the cosmological horizon. The existence of
a minimum uncertainty on the momentum may be related to the existence of a lower
bound for the mass of fields in \ads space [15].

Also in the case of EUP one can obtain the temperature of a black hole following the
same steps as above. It results
$$T\app{\D E\over2\p}={1\over4\p}\left[{1\over r_+}+{3r_+\over l_H^2}\right],
\eqno(7)$$
which is the standard temperature of an \ades \bh [16].

\section{3. Quantum mechanics in an \ades background}
One may suspect that the EUP can be derived from quantum mechanics defined on an
\ades background. The definition of \QM on \ades spacetime poses nontrivial problems
since, contrary to the case of flat spacetime, there is no privileged reference frame
like the one singled out by the \mi metric, and in fact very little can be found
on the subject in the literature [17].
However, it has been argued that a "natural" metric on \ades spacetime can be defined
by the choice of Beltrami (projective) \coo [17-19]. In particular, geodesic motion
takes place with constant velocity along straight lines in the spatial sections of
this metric [18].

Here we do not attempt the nontrivial task of defining \QM on an \ades background,
but simply give a heuristic construction of the commutation relations of single-particle
\QM in Beltrami coordinates, proceeding in analogy with the construction of the \pb of
\ades relativistic mechanics [19].

It is well known that \ades spacetime can be realized as a hyperboloid of equation
$\x_A^2=\pm\a^2$ embedded in 5-dimensional flat space, with coordinates $\x_A$ and
metric tensor $\y_{AB}={\rm diag}\ (1,-1,-1,-1,\mp1)$.

The isometries of \ades spacetime are generated by the \ades algebra. This can be
identified with the Lorentz algebra $so(1,4)$ (resp.\ $so(2,3)$) of the 5-dimensional
flat space, that leaves invariant the hyperboloid.
Its interpretation as 4-dimensional \ades algebra is obtained by splitting the
generators into Lorentz generators $J_\mn$ and translation generators
$p_\m=J_{4\m}/l_H$.
The \ades algebra can then be written as
$$\eqalignno{
&[J_\mn,J_{\r\s}]=i\left(\y_{\n\s}J_{\m\r}-\y_{\n\r}J_{\m\s}+\y_{\m\r}J_{\n\s}-
\y_{\m\s}J_{\n\r}\right),&\cr
&[J_\mn,p_\l]=i\left(\y_{\m\l}p_\n-\y_{\n\l}p_\m\right),\qquad
[p_\m,p_\n]=i\,{J_\mn\over l_H^2}\,.&(8)}$$

It is natural to identify the generators of translations $p_\m$ with the momentum
operators of a quantum theory defined on \ades spacetime.
Of course, the position operators $x_\m$ depend instead on the parametrization of the
hyperboloid. It is important to remark that, due to the curvature of spacetime, the
translation generators $p_\m$ will have nontrivial \cor with the positions $x_\m$.

The Lorentz subalgebra of the 4-dimensional \des algebra is identical to the
flat space Lorentz algebra, and hence its generators have the usual \cor
with the positions $x_\m$,
$$[J_{\mn},x_\l]=i\left(\y_{\m\l}x_\n-\y_{\n\l}x_\m\right),\eqno(9)$$
but the \cor of the translation generators with the position variables depend
instead on the specific choice of \coo on the hyperboloid.

The most natural parametrization of the hyperboloid is given by projective (Beltrami)
coordinates [17,18],
$$x_\m={\x_\m\over\x_4}.\eqno(10)$$
With this choice [19],
$$[x_\m,x_\n]=0,\qquad [x_\m,p_\n]=i\left(\y_\mn+{x_\m x_\n\over l_H^2}\right).
\eqno(11)$$
From the second equation and (3), the EUP (1) follows straightforwardly.

Notice that the form of the \cor strongly depends on the parametriz- ation of the
hyperboloid. The \coo usually adopted in general relativity, as cosmological or static
coordinates,
are not natural from a geometrical point of view, since they single out the time
coordinate. In some sense, they resemble the Rindler \coo on flat space.
The realizations of \QM generated by these parametrizations are of course not equivalent
to the one obtained here, and a discussion of their properties would require a deeper
investigation of \QM on an \ades background, that is beyond the scope of this letter.

\section{4. Extended generalized \up and triply special relativity}
It is natural to combine (1) and (2) to obtain a more general uncertainty principle,
which has been called EGUP [1,2],
$$\D x_i\D p_j\ge\d_{ij}\left[1+l_P^2(\D p_i)^2+{(\D x_i)^2\over l_H^2}\right].
\eqno(12)$$
In this case, if $l_P^2>0$ and $l_H^2>0$, a minimum measurable value both for length
and for momentum results, and, as in the case of GUP, the evaporation process of a
\bh stops when its radius reaches the minimal length.

The \up (12) recalls the commutation relations of triply special relativity (TSR).
TSR is a generalization of DSR to a curved background, based on a deformation
of the \ades algebra, such that both parameters $l_P$ and $l_H$ are observer
independent. As in flat space DSR, choosing different deformations, one can define a
variety of models exhibiting different properties [19,20].

In TSR models, the \cor of the \ades group generators maintain the form (8),
while the others \cor depend on the parametrization of the \ades hyperboloid and on
the specific
deformation of the algebra. We consider here the choice of [20], which is a direct
generalization of the choice of \coo adopted in the previous section for \ades
spacetime, and of the Snyder deformation discussed in sect.\ 2. In this model, the
\cor read
$$[x_\m,x_\n]=i\,l_P^2\,J_\mn,\qquad
[x_\m,p_\n]=i\left(\y_\mn+{x_\m x_\n\over l_H^2}+l_P^2\,p_\m p_\n+2{l_P\over l_H}
\,x_\m p_\n\right),\eqno(13)$$
from which the GEUP (10) follows in view of (3).

As usual, depending on the choice of the signs of $l_H^2$ and $l_P^2$ one obtains
different physical settings. For example, as mentioned above, in the anti-Snyder/\ads
case ($l_H^2>0$, $l_P^2>0$) a minimal value emerges both for length and momentum [2].
The temperature of a \bh is also modified and gives
$$T\app{\D E\over2\p}={r_+\over8\p l_P^2}\left[1-\sqrt{1-{4{l_P^2\over r_+^2}\left(
1-{r_+^2\over2l_H^2}\right)}}\ \right].\eqno(14)$$

\section{5. Conclusion}
We have obtained the EUP from geometric considerations on \ades spacetime.
Our derivation has also been extended to the case of the EGUP by using a deformation
of \ades relativity, known as triply special relativity [19,20].

The temperature of black holes in \ades \bg has sometimes been derived without
making recourse to the modifications of the \up [7].
In that case one obtains the corrections to the \sch temperature by introducing the
gravitational interaction as an external force on a flat background, and neglecting
the curvature of spacetime.
One may interpret that approach as equivalent to ours, but based on semi-Newtonian
gravity.

\beginref
\ref [1] B. Bolen and M. Cavagli\`a, \GRG{37}, 1255 (2005).
\ref [2] M.I. Park, \PL{B659}, 698 (2008).
\ref [3] G. Veneziano, Europhys. Lett. {\bf2}, 199 (1986).
\ref [4] M. Maggiore, \PL{B304}, 63 (1993).
\ref [5] M. Maggiore, \PL{B319}, 83 (1993).
\ref [6] C. Bambi and F.R. Urban, \CQG{25}, 095006 (2008).
\ref [7] F. Scardigli, \grq{0607010}.
\ref [8] F. Scardigli, \PL{B452}, 39 (1999);
R.J. Adler and D.I. Santiago, \MPL{A14}, 1371 (1999).
\ref [9] G. Amelino-Camelia, \PL{B510}, 255 (2001), \IJMP{D11}, 35 (2002).
\ref [10] J. Lukierski, H. Ruegg and W.J. Zakrzewski, \AoP{243}, 90 (1995);
J. Magueijo and L. Smolin, \PRL{88}, 190403 (2002);
F.J. Herranz, \PL{B543}, 89 (2002);
C. Heuson, \grq{0305015};
G. Amelino-Camelia, \IJMP{D12}, 1211 (2003).
\ref [11] H.S. Snyder, \PR{71}, 38 (1947).
\ref [12] S. Mignemi, \PL{B672}, 186 (2009).
\ref [13] S. Mignemi, in preparation.
\ref [14] R.J. Adler, P. Chen and D.I. Santiago, \GRG{33}, 2101 (2001);
M. Cavagli\`a, S. Das and R. Maartens, \CQG{20}, L205 (2003).
\ref [15] P. Breitenlohner and D.Z. Freedman, \AoP{144}, 249 (1982).
\ref [16] G.W. Gibbons and S.W. Hawking, \PR{D15}, 10 (1977);
S.W. Hawking and Page, \CMP{87}, 577 (1983).
\ref [17] R.L. Mallett and G.N. Fleming, \JMP{14}, 45 (1973).
\ref [18] H.Y. Guo, C.G. Huang, Z. Xu and B. Zhou, \PL{A331}, 1 (2004);
R. Aldrovandi R, J.P. Beltr\'an Almeida and J.P. Pereira, \CQG{24}, 1385 (2007).
\ref [19] S. Mignemi, \arx{0802.1129}; \arx{0807.2186}.
\ref [20] J. Kowalski-Glikman and L. Smolin, \PR{D70}, 065020 (2004).

\endref
\end